\documentclass[prl,twocolumn,showpacs,preprintnumbers,amsmath,amsfonts,amssymb,floatfix,aps,superscriptaddress]{revtex4-2}
\usepackage{graphicx}
\usepackage{xr}
\usepackage{enumitem}
\usepackage{amssymb}
\usepackage{amsmath}
\usepackage{amsfonts}
\usepackage{bm}
\usepackage{dsfont}
\usepackage{comment}
\usepackage{color}
\usepackage{relsize}
\usepackage{float}
\usepackage{bm}
\usepackage[normalem]{ulem}
\usepackage[nodisplayskipstretch]{setspace}
\usepackage{tikz}

\usepackage{hyperref}
\usepackage[export]{adjustbox}

\newcommand{\be}{\begin{equation}}
\newcommand{\ee}{\end{equation}}
\usepackage{units}

\begin{document}
\title{Tunable Field-Linked $s$-wave Interactions in Dipolar Fermi Mixtures}
%Andreas: I like the current title a lot! :-) The wording of the other two not not quite correct...
%Alternative titles
%\title{Field-Linked Molecules Enable Tunable $s$-wave Evaporation in Ultracold Dipolar Fermi Mixtures}
%\title{Engineering Repulsive Dipolar Shields and Feshbach-Like Resonances with Field-Linked Molecules}

\author{Jing-Lun Li}%
\email{Jinglun.Li@ist.ac.at}
\affiliation{Institute of Science and Technology Austria (ISTA), 
Am Campus 1, 3400 Klosterneuburg, Austria
}%

\author{Georgios M. Koutentakis}
\affiliation{Institute of Science and Technology Austria (ISTA), 
Am Campus 1, 3400 Klosterneuburg, Austria
}%

\author{Mateja Hrast}
\affiliation{Institute of Science and Technology Austria (ISTA), 
Am Campus 1, 3400 Klosterneuburg, Austria
}%

\author{Mikhail Lemeshko}
\affiliation{Institute of Science and Technology Austria (ISTA), 
Am Campus 1, 3400 Klosterneuburg, Austria
}%

\author{Andreas Schindewolf}
\email{andreas.schindewolf@tuwien.ac.at}
\affiliation{Vienna Center for Quantum Science and Technology,
Atominstitut, TU Wien, Stadionallee 2, 1020 Vienna, Austria}

\author{Ragheed Alhyder}%
\affiliation{Institute of Science and Technology Austria (ISTA), 
Am Campus 1, 3400 Klosterneuburg, Austria
}%

\begin{abstract}

Spin mixtures of degenerate fermions are a cornerstone of quantum many-body physics, enabling superfluidity, polarons, and rich spin dynamics through $s$-wave scattering resonances. Combining them with strong, long-range dipolar interactions provides highly flexible control schemes promising even more exotic quantum phases. Recently, microwave shielding gave access to spin-polarized degenerate samples of dipolar fermionic molecules, where tunable $p$-wave interactions were enabled by field-linked resonances available only by compromising the shielding. Here, we study the scattering properties of a fermionic dipolar spin mixture and show that a universal $s$-wave resonance is readily accessible without compromising the shielding.
We develop a universal description of the tunable $s$-wave interaction and weakly bound tetratomic states based on the microwave-field parameters.
The $s$-wave resonance paves the way to stable, controllable and strongly-interacting dipolar spin mixtures of deeply degenerate fermions and supports favorable conditions to reach this regime via evaporative cooling.
\end{abstract}

\maketitle

\section{Introduction}
Ultracold fermionic mixtures provide a powerful platform for studying strongly interacting quantum matter \cite{Inguscio:2007cma, bloch2008}. Their ability to scatter through $s$-wave collisions underpins efficient thermalization, evaporative cooling, and the emergence of superfluidity at low temperatures \cite{bederson_evaporative_1996}. The ability to tune interactions in atomic mixtures via magnetic Feshbach resonances \cite{Chin:2010} has led to landmark discoveries, including the BCS–BEC crossover and universal dynamics near unitarity \cite{zwerger2011}. Building on these advances, the incorporation of long-range dipolar interactions offers exciting prospects, from anisotropic superfluids \cite{Wenzel2018} to quantum spin liquids \cite{Savary_2017} and topological phases \cite{Kestner2011}. 
Ultracold dipolar fermionic spin mixtures have been realized using atomic species with large magnetic moments \cite{Chomaz:2023}. The dipole--dipole interaction is in this case, however, quite small and every state has a different dipole moment.
In contrast, ultracold molecules with electric dipole moments offer strong, long-range dipole--dipole interactions and a rich internal structure 
%AS: What do you want to cite here? What's the underlying logic? Review papers for sure... I added with Langen:2024 the most recent general review paper for dipolar molecules.
\cite{carr_cold_2009, Cooper2009,shi2010pra,Schmidt2022prr,Langen:2024}. This makes them ideal for exploring many-body physics and exotic quantum phases, 
%AS: Again, what is the rule for the citations? Only fermionic systems or all? More review based or concrete examples that can build up on your work?
\cite{bruun2008prl,iskin_ultracold_2007,Baranov:2012,wu_liquid_2016},
and the spin properties of the inter-molecular interaction offer 
%AS: I think SU(N) is a point that we should highlight, because our approach can certainly also be extended beyond 2 spin states.
 the potential for realizing SU($N$) symmetry \cite{Mukherjee:2025a,Mukherjee:2025b}. Therefore, it is highly desirable to explore the tunability of the interaction provided by such molecular systems.

However, dipolar gases also bring new challenges: on one side, the attractive part of the dipole--dipole interaction enhances inelastic collisions and induces collapse \cite{santos_bose-einstein_2000, Lushnikov2002};
%If these are too many citations, restrict it to Ospelkaus:2010 and Bause:2023
on the other side, ultracold molecules undergo inelastic processes in short-range collisions through chemical reactions or sticky complex formation \cite{Ospelkaus:2010,Mayle:2013,Christianen:2019,Liu:2020,Gregory:2020,Bause:2023}, 
complicating the route to deeply degenerate, stable mixtures. This limits the implementation of magnetic Feshbach resonances, the traditional tool for tuning the inter-molecular interactions, which are accessible only in specific molecular systems \cite{Park:2023}, and require access to short-range tetratomic states.
A powerful tool to overcome this challenge, the so-called microwave shielding, has recently emerged through microwave dressing of polar molecules.
%AS: I don't get the selection of paper that are cited here. There seems to be a disconnect from the content of the sentence and the citations. Either we cite some papers about microwave dressing of molecules, i.e., the dredecessors of the proper MW shielding plan, or we cite all kinds of shielding methods including the one that have nother to do with MW dressing. Both is fine, but it should be consistent.
%cite{Avdeenkov2006,Gorshkov2008,Micheli2010,Xie2020,matsuda_resonant_2020}
By coupling the rotational ground state to the first excited manifold with a blue-detuned circularly or elliptically polarized microwave field, an avoided crossing forms in the inter-molecular potential landscape, resulting in a field-induced barrier that dramatically suppresses inelastic losses at short range \cite{Cooper2009,Deng:2023, Huang:2012,Karman2018,Lassabliere2018,Anderegg:2021}. This further enables the tuning of the scattering properties through coupling to field-linked bound states that emerge in the interaction potential for suitable field parameters \cite{Avdeenkov:2002,Avdeenkov:2003,Avdeenkov:2004,Gorshkov2008,Cooper2009,Huang:2012,Lassabliere2018,matsuda_resonant_2020, Chen:2023,Chen:2024}. Microwave shielding has recently led to the creation of a stable degenerate dipolar Fermi gas \cite{Schindewolf:2022}, and 
% I am not against citing valtolina_dipolar_2020 but this would be the wrong context, as they did not use MW shielding. Or we need a different wording that is not exclusive to MW shielding.
%We should mention somewhere the first BEC, even if it does not directly lead to our work.
later this method was extended to dual-microwave shielding \cite{Deng:2025,Karman:2025}, which enabled the first creation of a BEC of dipolar molecules \cite{Bigagli:2024}.

%If you want less references you can cut one or two of the Avdeenkov papers. But it should at least keep the 2003 or 2004 paper, to demonstrate how the term 'field linked' came about. There are in principle also other papers, that we already cite and that could be listed here as well...

So far, microwave shielding has been restricted to gases composed of a single internal state. In the fermionic system, collisions occur primarily via $p$-wave channels.
%, limiting the tunability of the interaction and the evaporative cooling efficiency.
While enhanced $p$-wave interactions have recently been demonstrated by tuning to a field-linked resonance (FLR) \cite{Chen:2023}, this required a highly elliptical field polarization, which compromises the shielding efficiency and thereby the stability of the sample \cite{Karman:2019}. The fermionic systems with the lowest entropy ($T/T_F\simeq0.36$ \cite{Schindewolf:2022}, where $T_F$ is the Fermi temperature) have therefore been prepared far away from an FLR.

%As a result, the deep quantum degenerate regime in fermionic dipolar molecular gases remains elusive with the lowest temperature achieved being $T/T_F\simeq0.36$ \cite{Schindewolf:2022}.  

Here, we show that introducing a second spin state naturally provides tunable $s$-wave interactions in the experimentally accessible parameter regime with circular microwave polarization, i.e., without compromising the stability of the system through elliptical polarization. By solving the full coupled-channel scattering problem between two microwave-dressed molecules in different spin states, we find that the scattering length can be steered from strongly attractive to strongly repulsive by adjusting the microwave field parameters around a FLR.
%AS: I think we can cut here the following sentence, as I made the comparisson with Feshbach resonances earlier. However, if you want to stress that the universality is a feature that is not present with Feshbach resonances, we might want to mention the Feshbach resonance, again...
%This $s$-wave FLR plays an analogous role in turning interparticle interactions as the magnetic Feshbach resonances in atomic gases. However, it arises purely from the microwave-induced hybridization of rotational states.
Remarkably, this FLR exhibits universal behavior in terms of the field-linked parameters, persisting across different molecular species. This universality enables the prediction of $s$-wave scattering properties and weakly bound states via simple expressions. The combination of fermionic statistics, the abscence of ellipticity and deeper bound states provide favorable conditions for a stable, strongly-interacting and highly-tunable dipolar spin-mixture.
This universality enables the estimation of FLR positions and widths in a wide range of field parameters via a simple expression.

% This mechanism provides an unprecedented level of control over interactions in a long-lived, stable dipolar Fermi mixture. {\color{red} we make this paragraph more concise and highlight here also our key findings}
% Crucially, this tuning is achieved without sacrificing the field-linked repulsive barrier that suppresses dipolar loss. 

% \textcolor{blue}{[Overall the issue of double shielding is left hanging, shouldn't we add two words that it is negligible here due to the three-body centrifugal barrier? Could be also motivation for NaK.]}
%AS: I am not so confinced anymore that the centrifugal barrier will help a lot with the three-body recombination loss. We have seen in Schindewolf2022 that we reach with spin-polarized NaK the unitary regime in two-body collisions regardless of the centrifugal barrier, unless we detune the microwave significantly. If I translate the two-body interaction into the three-body collision, I think the suppression of three-body loss compared with the bosonic system might be relatively small. At least, it is not as crazy as in atomic 2-component Fermi gases.

\section{Results}
\begin{figure}
    \centering
     \includegraphics[width=1.0\linewidth]{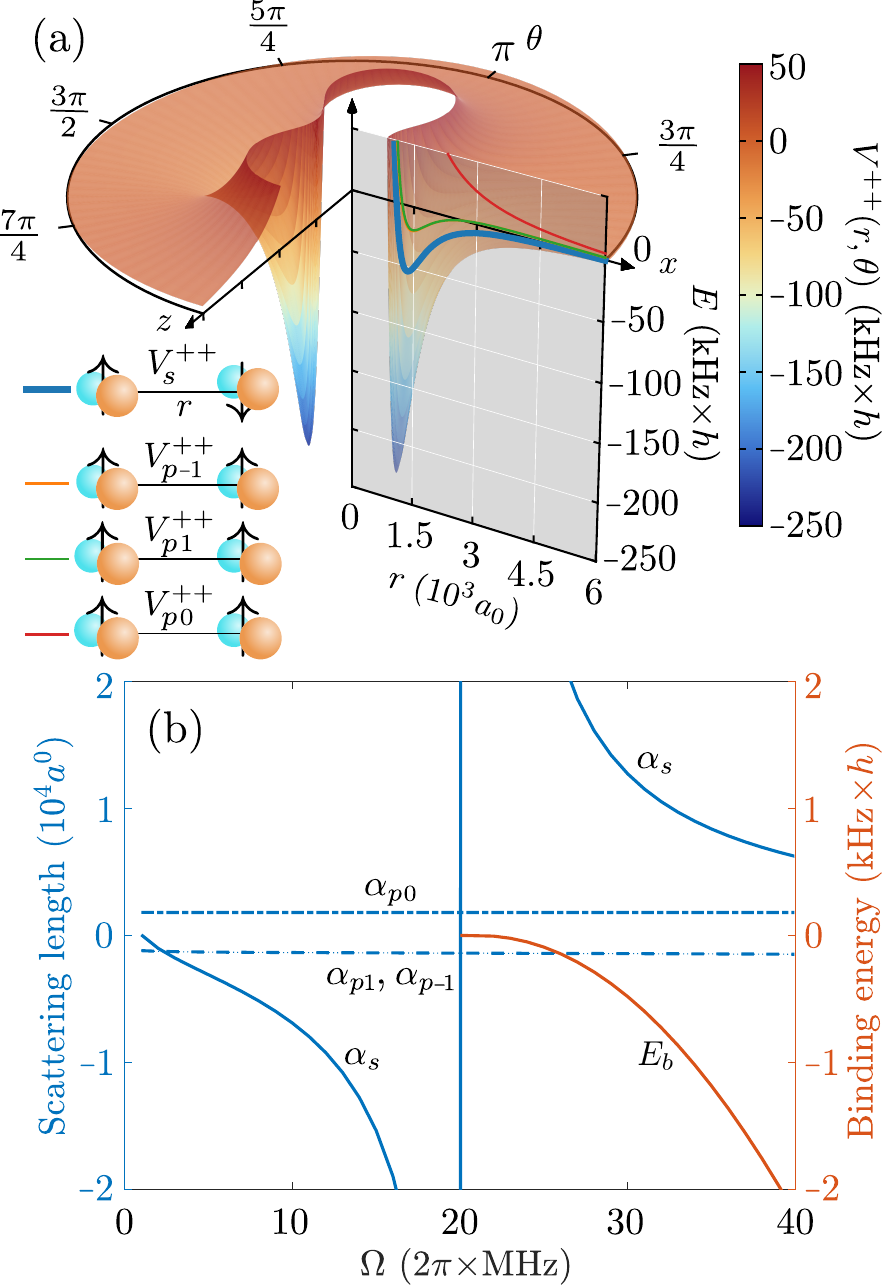}
    \caption{(a) Potential energy surface, $V^{++}(r,\theta)$ of the $|{++}\rangle$ collisional dressed state of microwave shielded molecules. The potential energy curves of $s$-wave $V_{s}^{++}(r)$ and $p$-wave $V_{pm_p}^{++}(r)$ channels, corresponding to unlike and like spins, respectively, are also provided. We note that the curves of $V_{p1}^{++}(r)$ and $V_{p-1}^{++}(r)$ are overlaped. In all cases ${}^{23}$Na$^{40}$K was considered with $\Omega = 2\pi \times \unit[21]{MHz}$ and $\delta=0$. (b) Inter- and intra-spin interactions characterised by the $s$-wave scattering length $\alpha_s$ and the energy-dependent $p$-wave scattering length $\alpha_{pm_p}(k)$, respectively, obtained at $\delta=0$. The binding energy $E_b$ of the $s$-wave weakly bound tetratomic state is also displayed. The energy-dependent $p$-wave scattering length is calculated at $E=21$ nK. In the unit, $a_0$ denotes the Bohr radius.
    %\andreasComment{In the text the $s$-wave potential is $V^{++}_s$, while in the figure, it is $V^{++}_{s0}$. I suggested to make it $V^{++}_s$ consistently.}
    }
    %AS: I know, pannel (a) already has a tone of information, but should we indicate the bound state in the V++00(r) potential?
    %AS: Normally I would try to avoid using $\times$ in the units and instead write for example $\Omega/(2\pi) (MHz)$, but it is not a big deal, if we don't change it. However, the spacings around $\times$ should be made consistent.
    \label{fig:asp}
\end{figure}
We construct the two-body problem starting from the microwave-dressed single–molecule Hamiltonian that couples the $J=0$ and $J=1$ rotational manifolds.  In the co-rotating frame the field produces four dressed states $\{|+\rangle,|-\rangle,|0\rangle,|\xi^{-}\rangle\}$ whose energies depend only on the control parameters. These consist of $\Omega$ the microwave Rabi frequency (coupling strength),
%AS: Can we call \Omega the microwave coupling strnegth? I find this to be a slighgtly more direct description of the physical property (but Rabi frequency is also not too bad...)
$\delta$ the detuning from the $J=1$ rotational state,
%AS: Could we turn \delta into a big \Delta to be more consistent with small and big lettering? Also a lot of shielding papers define \delta = \Delta/\Omega.
and $\xi$ the ellipticity of the field polarization (see Methods),
%AS: I refer here to the methode section to refer to the proper definition of the ellipticity.
which here is considered circular ($\xi=0$).
The molecules are prepared in the $|+\rangle$ state, leading to an anisotropic adiabatic interaction potential $V^{++}(r,\theta)$ with a strongly repulsive core (the shielding potential) in the two-molecule  $|{++}\rangle$ state, depicted in Fig.~\ref{fig:asp}(a). In the case of fermionic molecules, the internal spin state plays a crucial role in their scattering properties. In particular, two molecules with unlike spin states can collide in the $s$-wave channel of the shielding potential ($V^{++}_{s}$ in Fig.~\ref{fig:asp}(a)). In contrast, in the case of identical spin states, the lowest collisional channels have $p$-wave character ($V^{++}_{p0}$,$V^{++}_{p1}$ and $V^{++}_{p-1}$ ), see also Methods.

By tuning the microwave field parameters, one can introduce a FLR associated with a weakly bound tetratomic state (dimer state of two diatomic molecules). As shown in Fig.~\ref{fig:asp}(b), in the vicinity of the $s$-wave FLR, the scattering length characterising the inter-spin interaction is significantly enhanced while the intra-spin $p$-wave interaction remains almost unchanged.

In the following, we focus on fermionic ${}^{23}$Na${}^{40}$K molecules, which have been employed in previous experimental studies~\cite{Schindewolf:2022,Chen:2023,Chen:2024}, however, our results are universal and can be applied to other species \cite{Dutta:2025}. 
Our interaction Hamiltonian contains the electric dipole--dipole coupling within the rotating-wave approximation (RWA), van-der-Waals interactions, and an absorbing short-range boundary condition. Coupled-channel equations are solved, yielding elastic and inelastic scattering matrices from which all scattering observables follow (see Methods).
We present in the following a full characterisation of the FLRs, providing a simple expression for the scattering length, the position of the FLR, its width, and the weakly bound tetratomic state. Afterwards, we delve deeper into the scattering properties and the various enhancements FLRs can bring to experimental setups. 

\subsection{Universal characterisation of $s$-wave field-linked resonances and tetratomic molecules }
\begin{figure}
    \centering
    \includegraphics[width=1.0\linewidth]{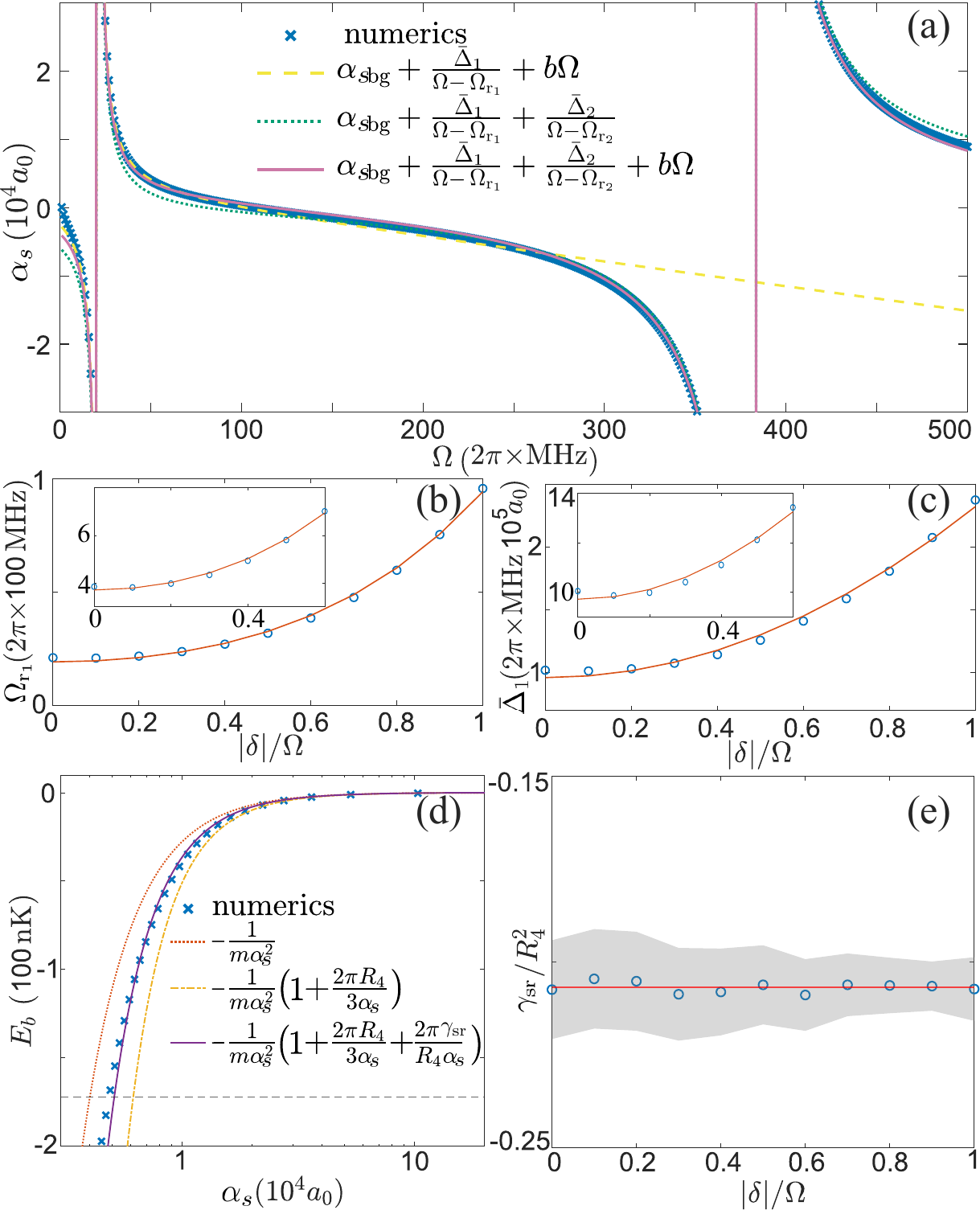}
    \caption{(a) scattering length versus microwave coupling strength at fixed $|\delta|/\Omega=0$ shows the first and second FLRs. The cross symbols represent numerical calculations, while the lines denote the fitting results. (b) and (c) the position $\Omega_{r_1}$ and effective width $\Delta_1$ of the first FLR increase with an increasing $|\delta|/\Omega$, respectively. The insets show the corresponding result for the second FLR. The solid lines in (b) and (c) the best fitting to Eqs (\ref{eq:flro}) and (\ref{eq:flrd}). (d) displays the binding energy of the weakly bound tetratomic state versus scattering length at $\delta/\Omega=0$, in the vicinity of the first FLR. The symbols represent the numerical calculation, while dotted and dashed lines indicate the universal formula to the order of $1/\alpha_s^2$ and $1/\alpha_s^3$ respectively. The solid line denotes the best fit to the $1/\alpha_s^3$ order, including the short-range correction parameter $\gamma_{\rm sr}$. (e) the scaled short-range parameters $\gamma_{\rm sr}/R_4^2$ versus $|\delta|/\Omega$ for the first FLR. The grey area indicates the $95\%$ confidence interval from the fitting. The red solid denotes the mean value of $\gamma_{\rm sr}/R_4^2=-0.21$.}
    \label{fig:scat}
\end{figure}

We study the FLR by computing the  $s$-wave scattering length at zero collisional energy limit while sweeping the microwave coupling strength $\Omega$. We find divergences reminiscent of Feshbach resonances in atomic physics (Fig.~\ref{fig:asp}(b)), which grant precise and tunable control over the $s$-wave interaction strength.

To characterise this field-linked resonance (FLR), we fit the calculated $s$-wave scattering length $\alpha_s$ (real part) to the following form:
\begin{equation} \label{eq:FR}
\alpha_s=\alpha_{s\rm bg}\left(1+\sum_i\frac{\Delta_i}{\Omega-\Omega_{\rm{r}_i}}\right)+b \, \Omega,
\end{equation}
where $\alpha_{s\rm bg}$ denotes the background scattering length, $\Omega_{\rm{r}_i}$ and $\Delta_i$ are the position and width of the $i$th FLR, respectively.
% \andreasComment{Strictly speaking $\Delta_i$ is not a 'resonance width'. A width should have the same unit as $\Omega$. If you would fit $a_{s,\text{bg}} \Delta_i$ or something similar in the numerator, then $\Delta_i$ would be a measure for the resonane width. Is there maybe a term that fits better then 'width'?}
In practice, we run a focused scan of $\Omega \in $ $2 \pi \times \unit[{[0,100]}]{MHz}$ for addressing the first FLR or a wide scan of $\Omega \in$ $ 2 \pi \times \unit[{[0,800]}]{MHz}$ for addressing the first two FLRs
fitting the resulting scattering length for $|\delta|/\Omega =0$, as exemplified in Fig.~\ref{fig:scat}(a). We extract the resonance positions $\Omega_{\rm{r}_i}$, background scattering length $\alpha_{s\rm bg}$, the linear coefficient $b$, and  the effective widths $\bar{\Delta}_i \equiv \alpha_{s\rm bg} \Delta_i $, defined to improve the numerical stability of the fit. We then proceed to fit the FLRs at various fixed $|\delta|/\Omega \in [0,1]$ to explore their control via field parameters (Fig.~\ref{fig:scat}(b-c)). 
% We got effective width term from PhysRevResearch.5.013174

For a given value of the detuning $|\delta|/\Omega$, the fitted background scattering length $\alpha_{s\rm bg}$ depends strongly on the scan range of $\Omega$ (see supplemental material \cite{sm}).
% \andreasComment{What das '(dataset)' mean here? Is the data for a given $|\delta|/\Omega$ a dataset? In that case the term should maybe be used earlier in the sentence?}
The observed variation in $\alpha_{s\rm bg}$ is remedied by incorporating the next-to-leading linear term ($b\,\Omega$ in Eq.~\eqref{eq:FR}) to achieve good fitting quality. We find that the inclusion of the linear correction helps to reproduce the resonance shape of the FLR. We attribute this feature to the existence of higher resonances for larger field strengths and to the fact that the interaction takes place on a single potential curve reshaped by the microwave field, leading to a varying background scattering length. In contrast, for magnetically tuned Feshbach resonances \cite{Chin:2010}, $\alpha_{s\rm bg}$ is fixed by the open channel parameters for a given atomic or molecular species.
%\andreasComment{I think this claim is too strong. Unless we can provide better arguments... What we see fro the 3 fits in Fig. 2(a) is that the second resonance has a huge effect on the first one and the same is probably the case for the following resonances as well. So, it is not necessarily a bad description of the 'background' scattering length, but certainly an incomplete picture, if the even stronger neighboring resonance is not part of the fit. You run into a similar problem with Feshbach resonances, e.g., in Cs-Cs, when you try to describe resonances that are overshadowed by stronger resonances. (that's where the product expression can be helpful). Now if there are other good arguments why the 'background scattering length' should have a linear dependence and if this linear dependence is always the same, regardless of how many resonances you try to fit, then we can make a more convincing argument...}
%In FLRs, however, the interaction takes place on a single potential curve reshaped by the microwave field, leading to a varying background scattering length  $\alpha_{s\rm bg}$. 
Note that the fit does not work for low $\Omega$ ($\unit[<10]{MHz}$), but in this regime there exists strong losses due to inefficient shielding. Crucially, the fitted positions and widths of the FLR remain stable if we change the scan range of $\Omega$ \cite{sm}.
% \andreasComment{Again, I am not sure what defines a dataset...}

 \begin{table*}[]
    \centering
     \caption{The fitted coefficients ($c_{\Omega_{\rm r}},c_{\Delta},\lambda_{\Omega_{\rm r}},\lambda_{\Delta}$) of position and width of the first (FLR1) and second (FLR2) field-linked scattering resonance. The values of $c_{\Omega_{\rm r}}$ and $c_{\Delta}$ are in $2\pi \times \unit[]{MHz}$ and $2\pi \times \unit[]{MHz}\, 10^4a_0$, respectively. To generalize the results to other species, the dimensionless coefficients are introduced by scaling $c_{\Omega_{\rm r}}$ and $c_{\Delta}$ as $ \tilde{c}_{\Omega_{\rm r}}=\hbar c_{\Omega_{\rm r}}/E_3$ and $\tilde{c}_{\Delta}=\hbar c_{\Delta}/E_3R_3$. The values of $ \tilde{c}_{\Omega_{\rm r}}$ and $\tilde{c}_{\Delta}$ are also listed (without error bar). Here $R_3=md^2/\hbar^2 4 \pi \epsilon_0$ and $E_3=\hbar^2/mR_3^2$ are the chracteristic length and energy scales of the dipole-dipole interaction, respectively \cite{Maykel:2017,Dutta:2025}.}
    \begin{tabular}{ccccccccc}
    \hline
    \hline
       &$c_{\Omega_{\rm r}}$&$\tilde{c}_{\Omega_{\rm r}}$&$c_{\Delta}$&$\tilde{c}_{\Delta}$&$\lambda_{\Omega_{\rm r}}$&$\lambda_{\Delta}$\\
       FLR1&19.11 (0.99)&5.76$\times$10$^{6}$&9.61(0.40)&2.20$\times$10$^{5}$&1.22 (0.07)&1.41(0.14) \\
       FLR2&369.5 (12.3)&1.11$\times$10$^{8}$&97.11(3.01)&2.23$\times$10$^{6}$&1.03(0.09)&1.01(0.20)\\
    \hline
    \hline
    \end{tabular}
    \label{tab:FR}
\end{table*}

%We observe in Figs. \ref{fig:scat}(b) and \ref{fig:scat}(c) that the position $\Omega_{\rm{r}_i}$ and effective width $\bar{\Delta}_i$ of both the first ($i=1$) and second ($i=2$) FLRs increase with $|\delta|/\Omega$.
The field dependence of the resonance position $\Omega_{\rm{r}_i}$ and effective width $\bar{\Delta}_i$ is governed by the long-range $-C_4/r^4$ potential, with $C_4=d^4m/[\hbar^2 1080(4\pi \epsilon_0)^2(1+(\delta/\Omega)^2)^2]$ for microwave dressed molecules \cite{Karman:2018pra,sm}, where $d$ denotes the molecule's permanent dipole moment and $\epsilon_0$ is the vacuum permittivity. In the asymptotic region $r \gg R_4 $, this potential dominates the inter-molecular interactions, and defines the characteristic length $R_4=(mC_4)^{1/2} \propto ( 1 + (|\delta|/\Omega)^2)^{-1}$ and energy $E_4=\hbar^2/mR_4^2 \propto ( 1 + (|\delta|/\Omega)^2)^2$ scales. For a pure  $-C_4/r^4$ potential, one finds $\Omega_{\rm{r}_i}\propto E_4$ and $\Delta_i\propto R_4E_4$ \cite{sm}, thus both quantities for the first ($i=1$) and second ($i=2$) FLRs increase with $|\delta|/\Omega$ as observed in Figs. \ref{fig:scat}(b) and \ref{fig:scat}(c). 
However, at intermediate and short-ranges, the $-C_4/r^4$ potential will be modified by the repulsive core of the adiabatic shielding potential and other forces. Hereafter, we use the term short-range specifically to describe the repulsive core of the adiabatic shielding potential (typically located at a few hundred $a_0$), which should not be confused with the regime of electron exchange (typically on the order of 10 $a_0$).

Consequently, to describe the increasing behaviour of $\Omega_{\rm{r}_i}$ and $\bar{\Delta}_i$, we derive the following simple relations \cite{sm}
\begin{align} \label{eq:flro}
 &\Omega_{\rm{r}_i}=c_{\Omega_{\rm r}} [1+\lambda_{\Omega_{\rm r}} (|\delta|/\Omega)^2]^2,
 \\\label{eq:flrd}
& \bar{\Delta}_{i}=c_{\Delta} [1+ \lambda_{\Delta} (|\delta|/\Omega)^2],
 \end{align}
which we use as fitting functions in Figs. \ref{fig:scat}(b) and \ref{fig:scat}(c). 
Here, $c_{\Omega_{\rm r}},\,c_{\Delta}$ are the scaling coefficients and $\lambda_{\Omega_{\rm r}}, \, \lambda_{\Delta} \neq 1$ represent short-range correction terms to the scaling, accounting for deviations from the $-C_4/r^4$ potential for $\Omega_{\rm{r}_i}, \, \bar{\Delta}_{i}$, respectively. %Equations (\ref{eq:flro}) and (\ref{eq:flrd}) recover characteristic scalings associated with $R_4$ and $E_4$ in the limit of $\lambda_{\Omega_{\rm r}}=\lambda_{\Delta}=1$.
%\andreasComment{Have you checked up to which distance the adiabatic potential is well described by $C_4/r^4$? It might only fit at long-long range and then turn into $1/r^3$ before reaching the intermediate range with the repulsive barrier? This is a quite important check, (should be stated in methods or supplement) if you draw conclusions based on the long-range potential shape...}
 
 % ragheed : Put this comment in the supplement
 % \textcolor{blue}{For the microwave shielded dipolar molecules in the $| + \rangle \otimes | + \rangle$ configuration/channel [or mention $V^{++}_{0,0}(r)$], we get $C_4=d^4m/1080(1+(\delta/\Omega)^2)^2$ \cite{Karman:2018pra}.}

% When the short-range correction factors are equal to unity ($\lambda_{\Omega_{\rm r}}=\lambda_{\Delta}=1$), observables %depend solely on characteristic scales of the long-range $-C_4/r^4$ potential, and the system exhibits universality, a %feature that becomes increasingly pronounced for higher resonances. Deviations of $\lambda_{\Omega_{\rm r}}$ and %$\lambda_{\Delta}$ from unity indicate non-universal contributions arising from the short-range physics.
% \andreasComment{Maybe we should move this paragraph should come one paragraph later? It is a bit weird to already discuss %the changes of the $\lambda$ before we refer to Tab.~1 for the first time.}

The scaling coefficients and short-range correction parameters are extracted by fitting Eqs.~\eqref{eq:flro} and \eqref{eq:flrd} and are summerized in Table \ref{tab:FR}.
%\andreasComment{Why Eqs. (17) and (18) and not (2) and (3)? Aren't these the same equations? If we provide these equations already in the main text, I see no reason, why they should be written out again in the supplement. A reference to the main text would be sufficient. Or is there a difference?}
We observe that both coefficients for the resonance position $c_{\Omega_{\rm r}}$ and the width $c_{\Delta}$ increase significantly from the first to the second FLR, indicating that the latter (occurring at a much higher microwave coupling strength) becomes considerably broader. 
For the first FLR, the short-range correction factors deviate from unity ($ \lambda_{\Omega_{\rm r}} = 1.22$ and $\lambda_{\Delta} = 1.41$), reflecting a non-negligible contribution from the short-range interaction. 
%\andreasComment{Just some thoughts about the wording "short-range": I understand that you apply here a methode that is typically used to describe bound states and scattering resonances that happen at much shorter length scales. In these systems, it makes sense to talk about a short-range correction etc. In our case everything is based on long-range dipole-dipole interactions. But the short-range regime is still relevant for the "reactive" collisions (that are sometimes called "reaching short-range"). So, I think, if we stick to "short-range" because of the analogy to other models, we should atleaset provide a disclaimer, where we state that this is not actually short-range, where the short-range tetramers would form...}
In contrast, the second FLR yields $ \lambda_{\Omega_{\rm r}}\simeq 1$ and $\lambda_{\Delta} \simeq 1$, indicating that short-range contributions are negligible. This trend arises because the larger Rabi frequency, at which the second FLR appears, pushes the repulsive barrier to smaller inter-molecular separations, thereby enhancing the role of the long-range $-C_4/r^4$ potential and limiting the short-range contribution.
%\andreasComment{I am a bit reluctant to call it a "weaker" barrier. That sounds like the shielding would get worse for higher Omega. However, if it wouldn't be for FLRs, the opposite would be the case (as long as Omega is significantly smaller the the roational constant). I understand that a C6 coefficient that is fitted to the repulsive potential would get smaller. Which is because the uncoupled dressed states intersect at a steeper angle with larger Omega... Maybe we can instead say that the repulsive barrier effectively gets steeper or something like that, which readers will not directly correlate with the shielding efficiency?}
 As a result, observables exhibit nearly universal scaling with $R_4$ and $E_4$ in the vicinity of the second FLR. Similarly, this universality is expected to persist for higher FLRs when $\Omega \ll 2B_{\rm rot}$, where $B_{\rm rot}$ denotes the molecule's rotational constant. 

A different universal behavior has recently been demonstrated in identical bosonic dipolar molecules~\cite{Dutta:2025}, where the properly rescaled scattering quantities are identical across different species for a given $|\delta|/\Omega$. We verify that such a species-based universality can be generalized to include two-component fermionic dipolar molecules \cite{sm}, complementing the field-based ($|\delta|/\Omega$-based) universality we find for the second (and higher) FLRs for a given molecular species. In practice, using the parameters from Table \ref{tab:FR}, our simple expressions (\ref{eq:flro}) and (\ref{eq:flrd}) provide an easy-to-use tool for estimating the scattering length in the vicinity of the first two FLRs of the two-component $^{23}$Na$^{40}$K molecular species.
%\andreasComment{Once more, shouldn't we refer to Eqs. (2) and (3)?}
This formalism is generally applicable to other two-component fermionic or identical bosonic dipolar molecules when incoparated with the spieces-based universality, by using the values of the recaled $\tilde{c}_{\Omega_{\rm r}}$ and $\tilde{c}_{\Delta}$ coefficients in Table \ref{tab:FR} in combination with the length ($R_3=md^2/\hbar^2 4 \pi \epsilon_0$) and energy ($E_3=\hbar^2/mR_3^2$) scales of dipole-dipole interaction.

In the following, we shall show that the field-linked tetratomic binding energy $E_b$ depends universally on the scattering length and $R_4$, up to energies on the order of $E_b \sim -E_4$, allowing our universal framework to be extended to estimate the field-linked tetratomic molecular binding energy.

To characterise the binding energy $E_b$ of a weakly bound field-linked tetratomic state, we employ the quantization function $F_4(E)$ of the long-range $-C_4/r^4$ potential \cite{Raab:2009}, supplemented by a short-range correction $\gamma_{\rm sr} mE/\hbar^2$ (see \cite{sm}). At $E=E_b$, the quantization function is related to the scattering length $\alpha_s$ via $\alpha_s=R_4/\tan[\pi \,F_4(E_b)]$. By solving for $E_b$ from and expanding the solution to third order in $1/\alpha$ we find
\begin{equation} \label{eq:Ebexpan}
    E_b=-\frac{\hbar^2}{m \alpha_s^2}\left( 1+ \frac{2\pi R_4}{3\alpha_s}+\frac{2\pi \gamma_{\rm sr}}{R_4\alpha_s} \right).
\end{equation}
%\textcolor{blue}{[Here we don't use the $\mathcal{O}$ notation since the next order has a $\ln$?]} 
%ragheed: refer to it in supplement, here it would be confusing 
In Eq.~\eqref{eq:Ebexpan}, the $1/\alpha_s^3$ terms supply, respectively, the leading long-range and short-range corrections to the universal relation $E_b=-\hbar^2/(m\alpha_s^2)$.  We note that for a pure $-C_4/r^4$ potential, the $R_4$ term has been introduced in Ref.~\cite{Idziaszek:2011}, while the $\gamma_{\rm sr}$ term is new and specific to the field-linked resonances. In Fig.~\ref{fig:scat}(d), we find that the universal relation is valid only when $|E_b|\ll E_4$, while for larger $E_b$, it underestimates the exact energy, while the long-range correction overestimates it. Including the short-range term restores quantitative agreement up to $|E_b|\approx E_4$ as is demonstrated in Fig.~\ref{fig:scat}(d). Remarkably, we find that the short-range parameter $\gamma_{\rm sr}$ normalized by $R_4^2$ is quite universal, for instance, $\gamma_{\rm sr}/R_4^2\approx-0.21$ for the first FLR, as is shown in  Fig.~\ref{fig:scat}(e). Hence, Eq.~\eqref{eq:Ebexpan} provides a practical and universal estimate of the field-linked binding energy estimated from $\alpha$ and $R_4$ in the regime $E_b\lesssim -E_4$. For higher FLRs, the universal value of $\gamma_{\rm sr}/R_4^2$ is expected to be much smaller.
\subsection{Collisional Landscape: Elastic Scattering and Loss Processes}

\begin{figure}
    \centering
    \includegraphics[width=1.05\linewidth]{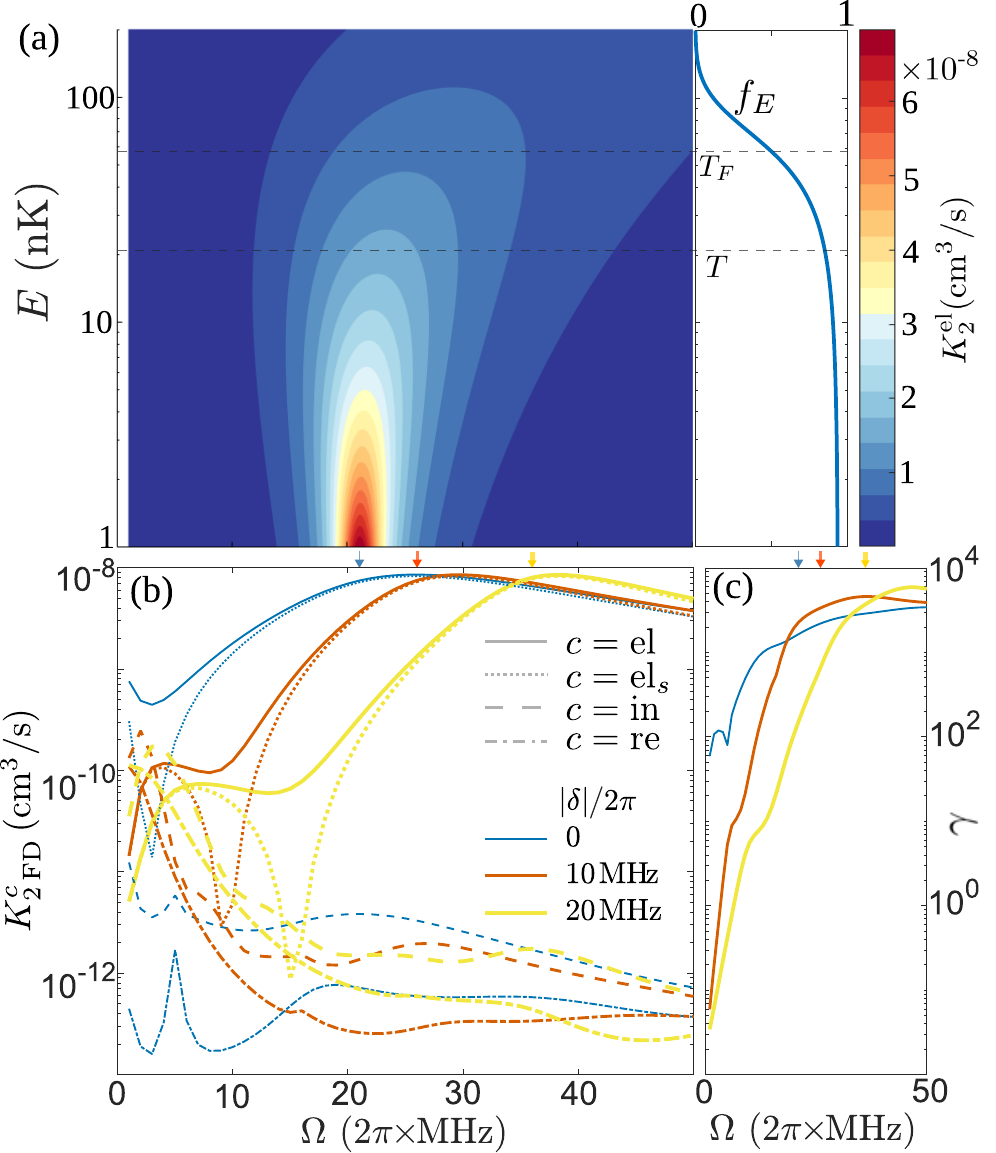}
    \caption{(a) Left panel: elastic scattering rate $K_2^\text{el}$ versus microwave coupling strength $\Omega$ and collisional energy $E$ at fixed detuning $\delta=0$.
 Right panel: Fermi-Dirac distribution $f_E$ with $T=21$ nK, $T_F=58$ nK and $\bar{\omega} = 2 \pi \times 60$ Hz, the typical experimental condition of Ref.~\cite{Schindewolf:2022}.
    (b) Thermal averaged elastic (solid lines), inelastic (dashed lines), and reactive (dash-dotted lines) scattering rate ($K_{2\rm FD}^\text{el}$, $K_{2\rm FD}^\text{in}$, and $K_{2\rm FD}^\text{re}$) considering $f_E$ from the right panel of (a). The dotted lines denote $K_{2\rm FD}^{\text{el}_s}$ the $s$-wave contribution to the elastic scattering rate. The detunings are $|\delta|/2\pi = 0, -10$ and $20 \text{ MHz}$ from thin to thick lines. (c) the ratio $\gamma=K_{2\rm FD}^{\rm el}/(K_{2\rm FD}^{\rm in}+K_{2\rm FD}^{\rm re})$ extracted from the result of (b). The arrows in (b) and (c) indicate the FLR positions at zero temperature.}
    \label{fig:rate}
\end{figure}

In the vicinity of the FLR, the scattering length, and consequently the elastic scattering rate $K^{\rm el}_2$, is significantly enhanced.  Figure~\ref{fig:rate}(a) shows that, at zero detuning ($\delta = 0$), tuning the microwave coupling strength to $\Omega\simeq 2\pi \times \unit[20]{MHz}$ boosts the elastic rate by nearly an order of magnitude over a broad collision‑energy interval $E \in [1,200]\,$nK.
This enhancement becomes narrower in $\Omega$ but grows in amplitude as $E$ is lowered, which is the desired behaviour for efficient evaporative cooling where colder molecules experience a higher probability of re‑thermalising collisions once the high‑energy tail is removed. 
Incorporating the Fermi-Dirac distribution in typical experimental conditions \cite{Schindewolf:2022}, the elastic scattering rate approaches the $s$-wave unitary limit $8.5\times10^{-9}$ cm$^3$/s in the $\Omega$ range of $2 \pi \times$ $\unit[{[20,\, 30]}]{MHz}$, as is shown in Fig.~\ref{fig:rate}(b).

The same calculation indicates that unfavorable inelastic and reactive scattering remain suppressed ( $<10^{-11}$ cm$^3$/s and $<10^{-12}$ cm$^3$/s, respectively), these are collisions that either scatter the molecules into other internal channels, or enter the regime where collisional complexes are formed within the repulsive core, respectively.
% %\andreasComment{I would like to add a brief explanation to the reactive scatter rate such as: "reactive scattering rates (i.e., collisions enter the regime where short-range tetramer states are formed)"...}
% The same calculation indicates that unfavorable inelastic and reactive scattering rates, collisions that scatter the molecules into other internal channels, and happen in the regime where tetramer states are formed within the repulsive core, respectively, 
% %\andreasComment{These collisions are in general not desirable. So maybe we can simply write "unfavorable inelastic and reactive scattering rates ..." and then cut the part about thermalization?}
% remain $<10^{-11}$ cm$^3$/s and $<10^{-12}$ cm$^3$/s, respectively, (see Methods for the definition of different scattering rates). 
This leads to the ratio $\gamma\equiv K_{2\rm FD}^{\rm el}/(K_{2\rm FD}^{\rm in}+K_{2\rm FD}^{\rm re})$ of the favorable to unfavorable collisions to be $\gamma>10^3$, as is shown Figure \ref{fig:rate}(c), which is suitable for evaporative cooling.
Close to the FLR, the elastic cross section is dominated by the isotropic $s$‑wave component (as shown in Fig.~\ref{fig:rate}(b)), which in addition to maximising $K_2^{\mathrm{el}}$, facilitates rapid cross‑dimensional thermalisation \cite{Bohn:2009,Wang:2021,Bigagli:2023}, beyond what is achievable with intra-spin $p$-wave interactions \cite{Schindewolf:2022}.
%a key advantage over $p$-wave interacting dipolar gases.
Away from the FLR, higher partial waves do contribute and can even dominate under some conditions. Increasing $|\delta|$ shifts the FLR to larger $\Omega$, where the suppression of losses is even stronger, leading to a larger $\gamma$. We note that the possible three-body recombination loss, which is encountered with NaCs \cite{Stevenson:2024,Yuan:2025}, can be excluded on the $\alpha_s<0$ side of the first FLR due to the absence of field-linked tetratomic (dimer) states. While a weakly-bound tetratomic state exists on the $\alpha_s>0$ side, enabling three-body recombination, the associated loss rate $L_3$ is expected to be suppressed at low temperatures ($L_3\propto E^{-1}$) by the three-body centrifugal barrier \cite{Esry:2001,DIncao:2018}.

A further practical benefit is that the onset of strong interactions occurs at comparatively low and experimentally available microwave field strengths, obviating the need for elliptically polarised dressing fields \cite{Chen:2023,Chen:2024} or extreme electric bias fields often required in static‑dipole schemes \cite{Mukherjee:2024}. This lower technical threshold simplifies experimental implementation and reduces residual heating typically present due to inefficient shielding \cite{Wang:2024}. In addition, high elastic scattering rates at reasonable microwave field strengths are key in view of microwave phase noise that induces effective single-body decay and that becomes more severe at larger field strengths \cite{Schindewolf:2022}. All these benefits facilitate evaporation to deeper degeneracy.
%while at the same time enabling deeper evaporation before intensity noise becomes limiting.
%\andreasComment{You mean intensity or phase noise? With phase noise we are certainly already fighting with the limits and going to higher MW field strengths will certainly make it difficult to keep the phase noise under control. At very large field strengths, intensity noise of the MW might also become relevant. For static electric field shielding noise is certainly also very critical. The evaporation benefits certainly also from better shielding (regardless of extra noise), while in this sentence it sounds a bit like noise is the main limiting factor for the evaporation. Or do you refer to the evaporation in my paper from 2022? In that case the time for the evaporation was indeed limited by phase noise (got improved by now, but can still become problematic again at at larger MW powers.)}
Since the elastic scattering rate and the density of states grow as the gas cools, the elastic collision rate actually increases as the sample approaches quantum degeneracy.  Consequently, the combination of a large $K_2^{\mathrm{el}}$, a high $\gamma$, and favourable isotropic $s$‑wave symmetry allows the system to cross into the degenerate regime at higher absolute temperatures and in shorter evaporation cycles than would be possible with $p$-wave interaction.

Finally, the enhanced scattering length pushes the system into the unitary‑limited interaction regime for $|\alpha_s|\gtrsim k_F^{-1}$, where $k_F$ is the Fermi momentum.  This increases the many‑body pairing gap and hence the superfluid transition temperature $T_c$ relative to weak‑coupling estimates \cite{Haussmann2007},
%\andreasComment{Can we add here a citation? Would Cooper:2009 be appropriate or do you have another suggestion?}
facilitating further cooling of the gas into the degenerate regime. Because the resonance occurs at $\Omega \ll 2B_{\mathrm{rot}}$, the microwave dressing leaves rotational coherence intact, minimising decoherence‑induced heating and making the FLR an attractive route towards efficient evaporation and strongly interacting dipolar Fermi superfluids.

\section{Discussion}
We have presented a comprehensive characterization of $s$-wave interactions and weakly bound tetratomic molecular states tuned by microwave field‑linked resonances of fermionic dipolar spin-mixtures. Starting from the microwave dressed Hamiltonian for two molecules with different spin states, we performed full coupled-channel scattering calculations. The extracted $s$-wave resonance properties largely follow universal scalings that depend solely on the microwave field parameters. We derive simple universal expressions for the resonance position, width and binding energy, and demonstrate that they remain accurate across a wide parameter range. Notably, the resonance parameters of higher‑order field‑linked resonances were shown to be increasingly universal, manifested by vanishing short‑range corrections. We refer to this universal behavior as the field-based universality, which is complemented by the species-based universality recently introduced in Ref.~\cite{Dutta:2025}. Consequently, the field-based universality is generally applicable to other two-component fermionic and identical bosonic species, and we verified this for two different molecular species.

Building on this microscopic control, we mapped the collisional landscape near the first FLR and demonstrated an interval in microwave coupling strength where the elastic rate coefficient approaches the unitary scattering limit, while inelastic and reactive channels remain three orders of magnitude smaller. In addition, the system is expected to be stable against three-body loss with respect to the bosonic case, an effect we will explore in a future study.
%\andreasComment{As mentioned in another comment of mine, I am not so convinced anymore that the centrifugal barrier will help a lot with the three-body recombination loss. We have seen in Schindewolf2022 that we reach with spin-polarized NaK the unitary regime in two-body collisions regardless of the centrifugal barrier, unless we detune the microwave significantly. If I translate the two-body interaction into the three-body collision, I think the suppression of three-body loss compared with the bosonic system might be relatively small. At least, it is not as crazy as in atomic 2-component Fermi gases. However, this does not means that we have to kill the argument. We could still write that we expect 3-body stability over a wide tuning range (up to the resonance). Or that we expect some suppression of three-body recombination compared to bosonic systems due to the centrifugal barrier (without quantifying it)...}
The enhanced isotropic $s$‑wave scattering in the vicinity of the FLR holds many evaporation-friendly features. In combination, they both promise to reach deep quantum degeneracy in ultracold dipolar fermionic molecules. This sets the stage for exploring anisotropic superfluidity, dipolar polarons, quantum magnetism, and novel topological phases.  
Beyond ultracold molecular gases, the universal scaling we uncover may inform further dipolar interaction control strategies in hybrid atom–ion systems \cite{Tomza2019} and long‑range Rydberg complexes \cite{shaffer_ultracold_2018}.

Taken together, these results elevate the microwave shielding of molecules from an experimental stabilisation technique to a versatile interaction‑engineering platform, similar to magnetic Feshbach resonances in ultracold atomic gases.  Future work will extend the present formalism to include many‑body effects, low‑dimensional confinement geometry, and time‑dependent dynamics, further broadening the horizons of strongly interacting dipolar quantum matter.

% \begin{itemize}
%     \item Better cross dimensional thermalization ($s$-wave dominated), maybe add $p$-wave on top.
%     \item Onset of strong interactions happens at lower Rabi frequencies, so no need for ellipticity.
%     \item Increased $T_C$ with $s$-wave scattering
%     \item Scattering increases with lower temperatures and less $T_F$
%     \item 3 body is small because of the large centrifugal barrier
%     \item For the first resonance, it should be high and experimentally accessible. No need for double shielding.
% \end{itemize}
\section{Methods}
Our Hamiltonian for two dipolar molecules in a microwave field reads
\begin{equation} \label{Ham}
    H=\sum_{i=1,2}\hat{h}_{i}+\hat{T}+V_{\rm vdW}+V_{\rm dd},
\end{equation}
where $\hat{h}_i=B_{\rm rot}\bm{J}^2-\bm{d}\cdot \bm{E}$ is the monomer hamiltonian (identical for both molecules), $\hat{T}$ denotes the kinetic operator for the relative motion between the two monomer centre-of-mass frames. Here, the $\bm{J}$ and $\bm{d}$ denote the rotational angular momentum and dipole moment vector of the monomer, respectively. The last two terms, $V_{\rm vdW}=-C_6/r^6$ and $V_{\rm dd}=\frac{d^2}{4 \pi \epsilon_0 r^3}\left[ \hat{\bm e}_{d_1} \cdot \hat{\bm e}_{d_2} -3 \left( \hat{\bm e}_{d_1} \cdot \hat{\bm e}_{r} \right)\left( \hat{\bm e}_{d_2} \cdot \hat{\bm e}_{r} \right) \right]$ describe the intermolecular van-der-Waals and dipole-dipole interactions, respectively, while $\epsilon_0$ denotes the vacuum permittivity. Here, $\hat{\bm e}_{d_1}$, $\hat{\bm e}_{d_2}$ and $\hat{\bm e}_{r}$ denote the unit vector of the dipoles $\bm{d}_1$ and $\bm{d}_2$ and the intermolecular position vector $\bm{r}$, respectively. We take the van-der-Waals coefficient $C_6$ as the sum of the induction $C_6^{\rm e}$ and dispersion $C_6^{\rm g-e}$ terms of Ref.~\cite{Lepers:2013}. 

We consider a microwave electric field $\bm{E}$
\begin{equation}
\bm{E}=Ee^{i(k_zz-\omega t)}(\cos \xi \hat{\bm{e}}_1+\sin \xi \hat{\bm{e}}_{-1})+\text{c.c.}
\end{equation}
at a frequency $\omega \approx 2 B_{\rm rot}/\hbar=\omega_{\rm m}$, near resonant with the $J=0\rightarrow J=1$ transition and with a ellipticity angle $\xi$, where $\hat{\bm{e}}_1$ and $\hat{\bm{e}}_{-1}$ are the unit vectors of circular and anticircular field, respectively. The microwave field couples the ground molecular rotational state $|J,m_J\rangle=|0,0\rangle$ with the excited superposition state $|\xi_{+}\rangle\equiv \cos \xi |1,1\rangle+\sin \xi |1,\text{-}1\rangle$ depending on the elliptical angle $\xi$. Two dark excited states, $|1,0\rangle$ and $|\xi_{-}\rangle\equiv \cos \xi |1,\text{-}1\rangle+\sin \xi |1,1\rangle$, to which the microwave field can not directly drive the transition, are also taken into account. Especially, these dark states become involved in two molecular collisions due to the dipole-dipole interaction $V_{\rm dd}$. Our model neglects higher excited rotational states with $J>1$.

In the frame co-rotating with the microwave field (defined by the unitary transformation $U=\exp (-i\omega\bm{J}^2t/2)$, the RWA can be employed to neglect the Hamiltonian terms that contain the fast time dependence factor $\exp(\pm i2\omega t)$. As a result, the system's Hamiltonian becomes static in the corresponding interaction picture defined by $H \to U^{\dagger} H U$. In particular, the monomer Hamiltonian yields two field-linked eigenstates $|+\rangle=u|0,0\rangle+v|\xi_{+}\rangle$ and  $|-\rangle=u|\xi_{+}\rangle-v|0,0\rangle$ with eigen energy $E_{\pm}=\hbar(\delta\pm \Omega_{\rm eff})$, where $u=-\sqrt{(1-\delta/\Omega_{\rm eff})/2}$ and $u=\sqrt{(1+\delta/\Omega_{\rm eff})/2}$. Here, $\delta=\omega_{\rm m}-\omega$ denote the detuning, $\Omega=2d E/\sqrt{3}\hbar$ is the bare Rabi requency and $\Omega_{\rm eff}=\sqrt{\delta^2+\Omega^2}$ the effective Rabi frequency. The energies of both dark states $|1,0\rangle$ and $|\xi_{-}\rangle$ are the bare detuning $\hbar\delta$ in the interaction picture. The validity of RWA requires the condition of $|\delta|\ll 2B_{\rm rot}/\hbar,\Omega \ll 2B_{\rm rot}/\hbar$.

The scattering problem of two field-linked molecules, governed by the Hamiltonian (\ref{Ham}), is numerically solved in the interaction picture using the coupled-channel Schr\"{o}dinger equation framework. The two-molecule internal channel states involved here are the product state in the monomer basis $|\nu\rangle\in\{|+\rangle,|-\rangle,|1,0\rangle, |\xi_-\rangle\}\otimes\{|+\rangle,|-\rangle,|1,0\rangle, |\xi_-\rangle\}$, yielding 16 total combinations. Through symmetrization, the 16 two-molecule basis states are classified into  10 symmetric basis $|\nu_{\rm s}\in \mathcal{S}\{|+\rangle,|-\rangle,|1,0\rangle, |\xi_-\rangle\}\otimes\{|+\rangle,|-\rangle,|1,0\rangle, |\xi_-\rangle\}$ and 6 antisymmetric basis $|\nu_{\rm a}\rangle \in \mathcal{A}\{|+\rangle,|-\rangle,|1,0\rangle, |\xi_-\rangle\}\otimes\{|+\rangle,|-\rangle,|1,0\rangle, |\xi_-\rangle\}$. Here, $\mathcal{S}=(1+P)/2$ and $\mathcal{A}=(1-P)/2$ denote the symmetric and antisymmetric permutation operators, defined by via $P$, of two molecules, respectively. The adiabatic Hamiltonian $H_{\rm ad}=\sum_{i=1,2}\hat{h}_{i}+V_{\rm vdW}+V_{\rm dd}$ defines the interaction potentials and couplings associated with these internal states. We note that $H_{\rm ad}$ commutes with the permutation operator and the couplings between the internal states in the symmetric sector $\{|\nu_{\rm s}\rangle\}$ and those in the antisymmetric sector $\{|\nu_{\rm a}\rangle\}$ are zero. Consequently, $H_{\rm ad}$ can be written as $H_{\rm ad}=H_{\rm ad}^{\rm s}\oplus H_{\rm ad}^{\rm a}$, a product sum of its projections in the symmetric sector $H_{\rm ad}^{\rm s}$ and in the antisymmetric sector $H_{\rm ad}^{\rm a}$.  Furthermore, three symmetric states ($|1,0\rangle|1,0\rangle, |\xi_-\rangle|\xi_-\rangle$ and $\mathcal{S}|1,0\rangle|\xi_-\rangle$) and one antisymmetric state ($\mathcal{A}|1,0\rangle|\xi_-\rangle$), which consist of only the dark states, are decoupled from others under RWA. These states can be removed from each basis set. As a result, the scattering problem can be studied either in the symmetric sector with 7 coupled $|\nu_{\rm s}\rangle$ channels or in the antisymmetric sector with 5 coupled $|\nu_{\rm a}\rangle$ channels, depending on the preparation of the incoming scattering state. In this work, we focus on the symmetric sector by considering two molecules prepared in the $|\nu_{s}^{\rm in}={++}\rangle\equiv|+\rangle|+\rangle$ scattering state. The adiabatic interaction potentials $V^{\nu_{\rm s}}(r,\theta,\phi)$ [$V^{\nu_{\rm a}}(r,\theta,\phi)$] are defined as the eigenvalues of $H_{\rm ad}^{\rm s}$ ($H_{\rm ad}^{\rm a}$) in the symmetric (antisymmetric) sectors. For instance, $V^{++}(r,\theta,\phi)$ denotes the adiabatic potential energy surface in the symmetric sector with a threshold the same as the channel energy of the $|{++}\rangle$ state. We note that without ellipticity ($\xi=0$ or $\pi$), potential surfaces $V^{\nu_{\rm s}}(r,\theta,\phi)$ [$V^{\nu_{\rm a}}(r,\theta,\phi)$] become independent of the azimuthal angle $\phi$. Accordingly, we omit $\phi$ from the notation in the main text. The anisotropic interaction potential $V^{\nu_{\rm s}}(r,\theta,\phi)$ couples different partial wave basis $|l m_l\rangle$ associated with the rotation of $\bm{r}=(r,\theta,\phi)$. Accordingly, one can define a combined basis set $\{|\nu_{s}lm_l\rangle\}$ that includes the partial wave. The Schr\"{o}dinger equation in $\{|\nu_{s}lm_l\rangle\}$ reads as a one-demensional coupled-channel equation of $r$
\begin{equation} \label{eq:cc}
-\frac{\hbar^2}{m}\frac{d^2}{dr^2}\psi_{\nu_{\rm s}lm_l}+\sum_{\nu'_{\rm s} l'm'_l}[H_{\rm ad}^{\rm s}+V_{\rm cen}]_{\nu_{\rm s} l m_l}^{\nu'_{\rm s} l'm'_l}\psi_{\nu'_{\rm s}l'm_l'}=E\psi_{\nu_{\rm s}lm_l},
\end{equation}
where $V_{\rm cen}=\hbar^2 l(l+1)/mr^2$ denotes the centrifugal interaction and $m$ is the monomer mass. The partial wave adiabatic potential curve $V^{\nu_{\rm s}}_{lm_l}(r)$ can be obtained by diagonalizing the matrix of resulting Hamiltonian $H_{\rm ad}^{\rm s}+V_{\rm cen}$. At large distance, the $V^{\nu_{\rm s}}_{lm_l}(r)$ is determinded by the dipole-dipole interaction $V_{\rm dd}$. This leads to a general aymptotic behavior $V^{\nu_{\rm s}}_{lm_l}(r)\xrightarrow[]{r\rightarrow \infty}C_{\nu_{\rm s}lm_l}/r^3+\hbar^2 l(l+1)/mr^2$, taking the diagonal element of the $V_{\rm dd}$. However, for $l = 0$ the leading term is $V^{\nu_{\rm s}}_{00}(r)\xrightarrow[]{r\rightarrow \infty}C_{\nu_{\rm s}00}/r^4$, arising from the second-order perturbation of $V_{\rm dd}$\cite{Karman:2018}. In the main text, the $s$- and $p$-wave partial wave adiabatic potential curves of $\nu_{\rm s}={++}$ are notated as $V_s^{++}(r)$ and $V_{pm_p}^{++}(r)$ with $m_p=0,\pm 1$, respectively. We also refer to $C_{++00}$ as $C_4$ in the main text.

The one-dimensional coupled-channel equation (\ref{eq:cc}) is solved by propagating the log-derivative matrix from a small intermolecular distance $r_{\rm in}$ with the absorbing boundary condition \cite{Wang:2015} to $r_{\rm out}$ at the asymptotic region, using the algorithm developed in Ref.~\cite{Manolopoulos:1986}. The scattering matrix $S$ is obtained by matching the log-derivative matrix at $r_{\rm out}$ to the asymptotic scattering wavefunction. From the $S$ matrix, one can define the partial elastic $K_{2,\nu^{\rm in}_{\rm s}lm_l}^{\rm el} (E)=\frac{2g\pi \hbar}{mk}|1-S_{\nu^{\rm in}_{\rm s}lm_l,\nu^{\rm in}_{\rm s}lm_l}|^2$, inelastic $K_{2,\nu^{\rm in}_{\rm s}lm_l}^{\rm in} (E)=\sum_{\nu'_{\rm s}l'm'_l\neq \nu^{\rm in}_{\rm s}lm_l} K_{2,\nu^{\rm in}_{\rm s}lm_l}^{\nu'_{\rm s}l'm'_l} (E)$ and reactive  $K_{2,\nu^{\rm in}_{\rm s}lm_l}^{\rm re} (E)=\frac{2g\pi \hbar}{mk}-\sum_{\nu'_{\rm s}l'm'_l}K_{2,\nu^{\rm in}_{\rm s}lm_l}^{\nu'_{\rm s}l'm'_l} (E)$ scattering rates for the incoming internal sate $|\nu^{\rm in}_{\rm s}\rangle$ at each partial wave $|lm_l\rangle$. Here, $K_{2,\nu^{\rm in}_{\rm s}lm_l}^{\nu'_{\rm s}l'm'_l} (E)=\frac{2g\pi \hbar}{mk}|S_{\nu^{\rm in}_{\rm s}lm_l,\nu'_{\rm s}l'm'_l}|^2$, $E$ is the collision energy to the threshold of $\nu_{\rm s}^{\rm in}={++}$ channel and $k=\sqrt{mE}/\hbar$. Note that the emergence of reactive scattering rate $K_{2,\nu^{\rm in}_{\rm s}lm_l}^{\rm re} (E)$ is due to the short-range absorbing boundary condition. The factor $g=2$ when the initial states of two molecules are identical (in all degrees of freedom, including spins), otherwise $g=1$. By collecting the contribution from all partial waves, we define also the total elastic $K_{2}^{\rm el} (E)=\sum_{l m_l}K_{2,\nu^{\rm in}_{\rm s}lm_l}^{\rm el} (E)+\sum_{lm_ll'm'_l\neq lm_l} K_{2,\nu^{\rm in}_{\rm s}lm_l}^{\nu_{\rm s}^{\rm in}l'm'_l} (E)$, inelastic $K_{2}^{\rm in} (E)= \sum_{\nu'_{\rm s}\neq \nu^{\rm in}_{\rm s}l m_ll'm'_l} K_{2,\nu^{\rm in}_{\rm s}lm_l}^{\nu'_{\rm s}l'm'_l} (E)$ and reactive $K_{2}^{\rm re} (E)=\sum_{lm_l}K_{2,\nu^{\rm in}_{\rm s}lm_l}^{\rm re} (E)$ scattering rates for the incoming internal sate $|\nu^{\rm in}_{\rm s}\rangle$. As $\nu_{\rm s}^{\rm in}={++}$ is fixed throughout this work, we remove it from the subscript for simplifying the notation. In this work, these scattering rates are calculated at various collision energies in $[1,1000]$ nK and are averaged according to the Fermi-Dirac distribution. The thermally averaged scattering rates are denoted with the subscript `FD'. We define the $s$-wave scattering length as $a_s=\lim_{k\rightarrow 0}\frac{1}{ik}\frac{1-S_{\nu_{\rm s}00,\nu_{\rm s}00}}{1-S_{\nu_{\rm s}00,\nu_{\rm s}00}}$ while for the $p$-wave we define an energy dependent scattering length $a_{pm_p}=\frac{1}{ik}\frac{1-S_{\nu_{\rm s}1m_p,\nu_{\rm s}1m_p}}{1-S_{\nu_{\rm s}1m_p,\nu_{\rm s}1m_p}}$ with $m_p=0,\pm 1$. We note that both $a_s$ and $a_{pm_p}$ are in principle complex values. We focus on the real part of the scattering lengths, and define $\alpha_s=\text{Re}(a_s)$ and  $\alpha_{pm_p}(k)=\text{Re}(a_{pm_p})(k)$. In the main text, we simply refer to $\alpha_s$ and $\alpha_{pm_p}$ as the $s$-wave scattering length and $p$-wave energy dependent scattering length, respectively.  

In addition to their internal rotational states, the molecules also possess internal spin states, denoted as $|\sigma_1\rangle$ and $|\sigma_2\rangle$ for the first and second components of the molecular spin mixture, respectively. For example, in case of $^{23}$Na$^{40}$K internal spin states correspond to specific configurations of the molecular nuclear spins such as $|\sigma_1\rangle=|S=0,M_S=0,m_{i_{\rm Na}}=3/2,m_{i_{\rm Na}}=-4\rangle$ and $|\sigma_2\rangle=|S=0,M_S=0,m_{i_{\rm Na}}=3/2,m_{i_{\rm Na}}=-3\rangle$. Under typical experimental conditions in the presence of a magnetic field, the nuclear spin of the molecules is decoupled from other degrees of freedom and is therefore treated as a spectator in the scattering process \cite{Will:2016,Karman:2018,Karman:2019,Karman:2025}. That is, including the molecular spin degree of freedom does not alter the overall adiabatic Hamiltonian $H_{\rm ad}$ of the microwave field-dressed molecules. Two molecules prepared in the state $|{++}\rangle$, but with different spin configurations, will collide on the same $V^{++}(r,\theta,\phi)$ potential surface, which couples identically with other $V^{\nu_{\rm s}}(r,\theta,\phi)$ potential surfaces associated with $H_{\rm ad}^{s}$. Nevertheless, under fermionic statistics, the spin state of the two molecules significantly affects their low-energy scattering properties. Given the above-defined mixtures, two molecules can have two intra-spin configurations, $|\sigma_1\sigma_1\rangle$ and $|\sigma_2\sigma_2\rangle$, and two inter-spin configurations $|\sigma_1\sigma_2\rangle^{\rm s}=(|\sigma_1\sigma_2\rangle+|\sigma_2\sigma_1\rangle)/\sqrt{2}$ and $|\sigma_2\sigma_2\rangle^{\rm a}=(|\sigma_1\sigma_2\rangle-|\sigma_2\sigma_1\rangle)/\sqrt{2}$. The two intra-spin configurations and inter-spin $|\sigma_1\sigma_2\rangle^{\rm s}$ state are symmetric, while the inter-spin $|\sigma_2\sigma_2\rangle^{\rm a}$ is antisymmetric. For two fermionic molecules in the $|{++}\rangle$ rotational state, the antisymmetry of the total wave function requires the odd partial waves $l=1,3,5,\cdots$ for the symmetric spin state, while even partial waves $l=0,2,4,\cdots$ for the antisymmetric spin state. As a result, the lowest collisional channel is $s$-wave for inter-spin scattering in $|\sigma_2\sigma_2\rangle^{\rm a}$, whereas it is $p$-wave for intra-spin scatterings and for inter-spin scattering in $|\sigma_2\sigma_2\rangle^{\rm s}$, where a centrifugal barrier must be overcome.

\section*{Data Availability}
The data that support the findings of this study are available from the corresponding authors upon reasonable request.

\bibliographystyle{unsrt}
%\bibliography{library.bib}

\section*{acknowledgments}
% We thank ... for fruitful discussions.
J. Li thanks Gaoren Wang for valuable discussions on the absorbing boundary condition.
G. M. K. thanks P. Giannakeas for fruitful discussions during the initial stages of this study. 
G. M. K. was funded by the Austrian Science Fund (FWF) [10.55776/F1004]. 
R. A. received funding from the Austrian Academy of Science ÖWA grant No. PR1029OEAW03.

\section*{Author contributions}
All authors contributed to the development of the research, the writing of the manuscript, and the interpretation of the results.  The theoretical calculations and its numerical implementation were performed by J.~L.~and G.~M.~K. R.~A.~and A.~S.~proposed the project and contributed to the interpretation of the results and to the writing of the manuscript.

\section*{Competing interests}
The authors declare no competing interests.

\section*{Materials and Correspondence}
Correspondence and requests for materials should be addressed to J. Li or R. Alhyder.
\newpage
\section{Supplemental Material}
 \subsection{Universality and comparison to other species}

\begin{figure}
     \centering
     \includegraphics[width=1.0\linewidth]{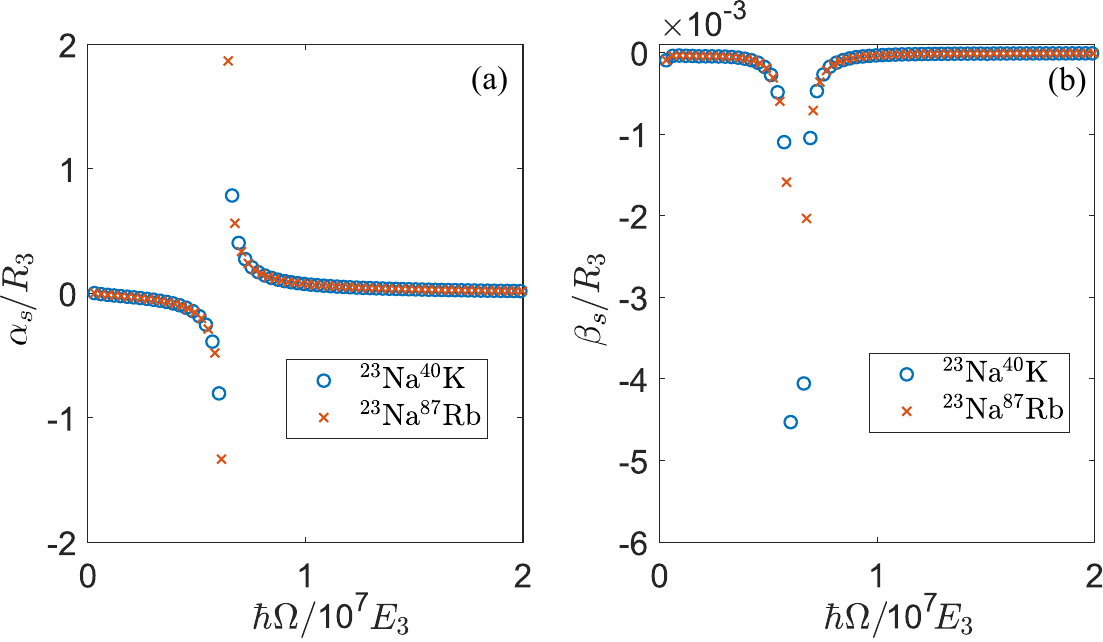}
     \caption{Comparison of real $\alpha_s$ (a) and imaginary $\beta_s$ (b) part of the scattering length in two-component $^{23}$Na$^{40}$K and single-component $^{23}$Na$^{87}$Rb gases. The detuning $\delta$ is considered to be zero. }
     \label{fig:uni}
 \end{figure}

 It has been demonstrated that the scattering property of microwave-linked identical bosonic dipolar molecules is universal in units of the chracteristic length $R_3=md^2/\hbar^2 4 \pi \epsilon_0$ and energy $E_3=\hbar^2/mR_3^2$ scales of the dipole-dipole interaction \cite{Dutta:2025}. The validity range of this universality is expected to extend to two-component fermionic molecules, since in both cases the collisions occur on the same $V^{++}(r,\theta,\phi)$ potential surface via the $s$-wave. We confirm the generalisation of this universality by comparing the scattering length of $^{23}$Na$^{40}$K in unlike spin states to that of $^{23}$Na$^{87}$Rb in identical spin states, scaled by $R_3$ and plotted against $\hbar\Omega/E_3$, as is shown in Fig.~\ref{fig:uni}. This establishes that the derived analytical expressions [Eqs.~\eqref{eq:flro} and \eqref{eq:flrd}] are generally applicable to other two-component fermionic as well as identical bosonic dipolar gases. Notably, they can serve as an easy tool for estimating the interaction strength and weakly bound state (when incorporated with Eq.~\ref{eq:Ebexpan}) in microwave field-dressed dipolar molecular gases. 

\subsection{Scaling formulas for $\Omega_{\rm{r}_i}$ and $\bar{\Delta}_i$}

Although multi-channel scattering induces appreciable shifts, the characteristic scales of field-linked resonances (FLRs) can be extracted by restricting the analysis to the $s$-wave channel of the incoming $|{++}\rangle$ state. The interaction is, to a good approximation, captured by the effective potential ~\cite{Deng:2023}:  
\begin{equation}
\begin{split}
    &V_{\rm eff}(\mathbf r)=\frac{\sqrt{4\pi}\,A}{r^{3}}Y_{2,0}(\theta,\phi) \\
    &~+\frac{\sqrt{4\pi}\,B}{r^{6}}\!\left[7Y_{0,0}(\theta,\phi)-\frac{Y_{2,0}(\theta,\phi)}{\sqrt5}-\frac{2Y_{4,0}(\theta,\phi)}{3}\right],
\end{split}
\end{equation}
with couplings
\begin{align}
    A&=\sqrt{\frac15}\,\frac{d^{2}}{24\pi\epsilon_{0}}\left[1+\left(\frac{|\delta|}{\Omega}\right)^{2}\right]^{-1}, \label{C4taoshi}\\
    B&=\frac{d^{4}}{1120\pi^{2}\epsilon_{0}^{2}\Omega}\left[1+\left(\frac{|\delta|}{\Omega}\right)^{2}\right]^{-3/2}.
    \label{eq:C6taoshi}
\end{align}

Projecting $V_{\rm eff}$ onto the $s$-wave subspace,
we found that only the $1/r^{6}$ term survives at the leading order $\langle 00|V_{\rm eff}|00\rangle$. However, the $1/r^{3}$ term can contribute at the second order via the coupling of $s$ and $d$ channels, $\langle 00|V_{\rm eff}|20\rangle\langle 20|V_{\rm eff}|00\rangle/V_{\rm cen}$, resulting in a $1/r^4$ interaction \cite{Karman:2018pra}. This leads to the effective $s$-wave potential
\begin{equation}
    V^{\rm eff}_{00}(r)\approx \frac{7B}{r^{6}}-\frac{C_4}{r^{4}},
    \label{eq:potential_tao_shi}
\end{equation}
with $C_4=mA^2/6\hbar^2=d^4m/[\hbar^2 1080(4\pi \epsilon_0)^2(1+(\delta/\Omega)^2)^2$, from which we define the long-range chracteristic length 
\begin{equation}
R_{4}=d^2m/[\hbar \sqrt{1080} (4\pi \epsilon_0) (1+(\delta/\Omega)^2)]
\end{equation}
and energy
\begin{equation}
E_{4}=1080(4\pi \epsilon_0)^2[1+(\delta/\Omega)^2]^2\hbar^{4}/(m^2d^4)
\end{equation}
scales. As the scattering properties of the molecules are largely determined by the long-range $-C_4/r^{4}$ interaction, one expects that $\Omega_{\rm r_i} \propto E_4$ and $\bar{\Delta}_i \propto R_4 E_4$ according to dimensional analysis, indicating $\Omega_{\rm r_i}\propto [1+(\delta/\Omega)^2]^2$ and $\bar{\Delta}_i \propto 1+(\delta/\Omega)^2$. At short and intermediate range, additional interactions such as the $7B/r^{6}$ term become significant, modifying the overall scaling of the dominant $-C_4/r^{4}$ potential. To describe such modification, we introduce a correction parameter $\lambda_{\Omega_{\rm r}}$  
\begin{equation}
\Omega_{\rm r_i}\propto [1+\lambda_{\Omega_{\rm r}}(\delta/\Omega)^2]^2
\end{equation}
for $\Omega_{\rm r_i}$, and $\lambda_{\Delta}$
\begin{equation}
\bar{\Delta}_i \propto 1+\lambda_{\Delta}(\delta/\Omega)^2
\end{equation}
for $\bar{\Delta}_i$. 

In the following, we derive the short-range correction parameters according to the effective $s$-wave interaction potential \ref{eq:potential_tao_shi}, using the first FLR as an example.
We render dimensionless with
$R_{4}$,
$E_{4}$ and
$\chi=7B/(R_{4}^{6}E_{4})$,
yielding the radial Schrödinger equation
\begin{equation}
    \left[-\frac{{\rm d}^{2}}{{\rm d}r^{2}}-\frac1{r^{4}}+\frac{\chi}{r^{6}}\right]\psi(r)=\epsilon\,\psi(r).
\end{equation}

The log-derivative calculation show the first FLR occurs at
$\chi=\chi_{1}=0.11$, giving
\begin{eqnarray}\label{eq:resonance_tao_shi}
    \Omega_{\rm{r}_1}&=&\frac{1}{0.11}\,
    \frac{1866240\,\pi^{2}\epsilon_{0}^{2}\hbar^{6}}{5d^{4}m^{3}} \\ \notag
    &\times&\left\{1+\frac{5}{2}\left(\frac{|\delta|}{\Omega}\right)^{2}+\frac{15}{8}\left(\frac{|\delta|}{\Omega}\right)^{4}+O\left[\left(\frac{|\delta|}{\Omega}\right)^{6}\right]\right\}. \\ \notag
    &\approx&\frac{1}{0.11}\,
    \frac{1866240\,\pi^{2}\epsilon_{0}^{2}\hbar^{6}}{5d^{4}m^{3}}\left[1+\lambda_{\Omega_{\rm r}}\left(\frac{|\delta|}{\Omega}\right)^{2}\right]^2
\end{eqnarray}
with $\lambda_{\Omega_{\rm r}}=1.25$. The estimated $\lambda_{\Omega_{\rm r}}=1.25$ is in good agreement with our numerical result $\lambda_{\Omega_{\rm r}}=1.22$.
Near resonance the $s$-wave scattering length exhibits a simple pole,
\begin{equation}
    \frac{\alpha}{R_{4}}\simeq\frac{\alpha_{\mathrm{bg}}}{R_{4}}+
    \frac{\Delta\chi_{i}}{\chi-\chi_{i}},
\end{equation}
where $\Delta\chi_{i}$ is obtained numerically (e.g.\ via log-derivative analysis).
Re-expressing in terms of $\Omega$ gives
\begin{equation}
    \alpha
    =\left(\alpha_{\mathrm{bg}}-\frac{\Delta\chi_{1}}{0.11}\right)R_{4}
    -\frac{\overbrace{(\Delta\chi_{1}/0.11)\Omega_{\rm{r}_1}R_{4}}^{=\bar{\Delta}_1}}{\Omega-\Omega_{\rm{r}_1}}
\end{equation}
for the first FLR. The corresponding effective width reads
\begin{equation}
   \bar{\Delta}_{1}\approx\frac{\Delta\chi_{1}}{0.11}\,
    \frac{2592\sqrt{30}\,\pi\epsilon_{0}\hbar^{3}}{d^{2}m^{2}\Omega_{\rm{r}_i}}
    \left[1+\lambda_{\Delta}\left(\frac{|\delta|}{\Omega}\right)^{2}\right].
    \label{eq:width_tao_shi}
\end{equation}
with $\lambda_{\Delta}=1.5$, which also agree well with our numerical result $\lambda_{\Delta}=1.41$. For higher FLRs, the decrease of $B$ with increasing $\Omega$ in Eq. (\ref{eq:potential_tao_shi}) implies that the short-range correction will become less significant at higher $\Omega$. Consistently, our numerical result shows $\lambda_{\Omega_i} \approx \lambda_{\Delta_i} \approx 1$ for the second FLR.

Equations~\eqref{eq:resonance_tao_shi} and \eqref{eq:width_tao_shi} are valid only in the long-range regime
$r^{3}>d^{2}/(4\pi\epsilon_{0}\hbar\Omega)$ assumed in Ref.~\cite{Deng:2023} within our perturbative approach. In reality,  additional terms enter the effective potential, which may explain the small deviations of the numerical value of $\lambda_{\Omega_{\rm r}}$ and $\lambda_{\Delta}$ from our estimations. 

\subsection{Derivation of the expansion formula for $E_b$}
 \begin{table*}[]
    \centering
     \caption{The fitting parameters of the first and second FLRs at $|\delta|/\Omega=0$.}
    \begin{tabular}{cccccccc}
    \hline
    \hline
       &scan range&$\alpha_{s\rm bg}$&$b$&$\Omega_{\rm r_1}/2\pi$&$\bar{\Delta}_1/2\pi$&$\Omega_{\rm r_2}/2\pi$&$\bar{\Delta}_2/2\pi$\\
       model&[$2\pi$MHz]&[$a_0$]&[$a_0$/MHz]&[MHz]&[$10^5a_0$MHz]&[MHz]&[$10^5a_0$MHz] \\
       (i)&[0,100]&2352&-35.37&20.97&1.027&-&-  \\
       (ii)&[0,800]&1584&-&20.97&1.028&383.5&10.03 \\
       (iii)&[0,800]&3686&-8.373&20.97&1.027&383.5&10.04 \\
    \hline
    \hline
    \end{tabular}
    \label{tab:fit}
\end{table*}
To get the expansion formula (\ref{eq:Ebexpan}) for $E_b$, we employ the quantization function $F_4(E)$ for $1/r^4$ potential \cite{Raab:2009}:
\begin{equation}
F_4(E)=A(E)F^{\rm low}_4(E)+[1-A(E)]F_4^{\rm high}(E)+F_4^{\rm sr}(E),
\end{equation}
where $F^{\rm low}_4(E)$ and $F_4^{\rm high}(E)$ are the low-$\kappa$ and high-$\kappa$ limit forms of the quantization function of the potential tail of $-C^4/r^4$, with $\kappa=\sqrt{-mE}/\hbar$. They read as
\begin{equation}
F^{\rm low}_4(E)=\frac{R_4 \kappa}{\pi}-\frac{(d_4 \kappa)^2}{2 \pi}
\end{equation}
and 
\begin{eqnarray}
F^{\rm high}_4(E)=-\frac{1}{4}+\frac{\Gamma(3/4)(\kappa R_4)^{1/2}}{\Gamma(5/4)2\sqrt{\pi}}+\frac{D_1/2\pi}{(\kappa R_4)^{1/2}} \\ \notag
+\frac{D_3/2\pi}{(\kappa R_4)^{3/2}} +\frac{D_5/2\pi}{(\kappa R_4)^{5/2}}+\frac{D_7/2\pi}{(\kappa R_4)^{7/2}},
\end{eqnarray}
where $d_4=\sqrt{2\pi/3}R_4$ are referred to as the threshold length and effective length, respectively. The $D_j$ denote the coefficients of the $j/2$-th order term of $(1/\kappa R_4)$ in the high-order WKB approach, which is documented in Ref.~\cite{Raab:2009}. The interpolation function $A(E)$ is selected as
\begin{equation}
A(E)=\frac{1}{1+(\kappa B_6)^6+(\kappa B_7)^7},
\end{equation}
with the fitted parameters $B_6=1.622576 R_4$ and $B_7=1.338059 R_4$. The short-range quantization function $F_4^{\rm sr}$ is a smooth analytic function that vanishes at threshold $E=0$
\begin{equation}
F_4^{\rm sr}(E)=\gamma_{\rm sr} E+\gamma_{\rm sr2} E^2+O(E^3).
\end{equation}
The scattering length is related to $F_4(E=E_b)$ via \cite{Raab:2009}
\begin{equation}
\alpha_s=\frac{R_4}{\tan[F_4(E_b)\pi]}.
\end{equation}
Next, we expand $R_4/\tan[F_4(E_b)\pi]$ in terms of $\kappa$
\begin{equation}
\frac{R_4}{\tan[F_4(E_b)\pi]}=1/\kappa+c_0+c_1 \kappa+c_2 \kappa^2+\cdots,
\end{equation}
where $c_0=(d_4^2+2 \pi \gamma_{\rm sr})/2R_4$, $c_1=(-4 R_4^4 + 3 d_4^4 + 12 d_4^2 \pi \gamma_{\rm sr} + 12 \pi^2 \gamma_{\rm sr}^2)/12 R_4^2$. The parameter $\gamma_{\rm sr 2}$ enters at the order of $c_2 \kappa^2$ in the expansion, whereas the contributions from the parameters in $F^{\rm high}_4(E)$ only arise beginning at the $c_5 \kappa^5$ order. By solving
\begin{equation}
\alpha_s=1/\kappa+c_0+c_1 \kappa+c_2 \kappa^2
\end{equation}
and expanding the solution in terms of $1/\alpha_s$, we get
% \begin{equation}
% E_b=-\frac{\hbar^2}{m}\left(\frac{1}{a^2}+\frac{2c_0}{a^3}+\frac{3c_0^2+2c_1}{a^4} \\
% +\frac{4 c_0^3+8c_0c_1+2c_2}{a^5}+\cdots \right).    
% \end{equation}
\begin{align}
E_b &=-\frac{\hbar^2 \kappa^2}{m} \\
&= -\frac{\hbar^2}{m} \left( 
\frac{1}{\alpha_s^2} + \frac{2c_0}{\alpha_s^3} + \frac{3c_0^2 + 2c_1}{\alpha_s^4} \right. \notag \\
&\left. + \frac{4c_0^3 + 8c_0 c_1 + 2c_2}{\alpha_s^5} + \cdots \right)
\end{align}
This result shows that $c_2$ first appears in the $E_b$ at the order of $1/\alpha_s^5$. In principle, one can truncate the expansion of $E_b$ at this order to exclude the contribution of $c_2$, thereby eliminating all terms involving $\gamma_{\rm sr 2}$. In practice, we truncate at the order of $1/\alpha_s^4$ and get
\begin{equation}
E_b=-\frac{\hbar^2}{m}\left[\frac{1}{\alpha_s^2}+\frac{b_2}{R_4}\frac{1}{\alpha_s^3}+\mathcal{O}\left(\frac{1}{\alpha_s^4}\right)\right]
\end{equation}
with $b_2=d_4^2+2\pi \gamma_{\rm sr}$, which is the Eq.~\eqref{eq:Ebexpan} in the main text. We note that the same expansion can be obtained in this order by applying the quantum defect theory to the analytical solution of the $-C_4/r^4$ potential \cite{Gao:2013}.
\subsection{Details of numerical simulation}
In our numerical simulation, we propagate the log-derivative matrix $\mathbf{Y}$ from an initial radius of $r_{\rm in}= 15$ $a_0$ to a final radius $r_{\rm out}= 2000$ $R_4$, where $R_4$ depends on microwave field parameters. Here, $a_0$ denotes the Bohr radius. At $r_{\rm in}$, the full absorbing boundary condition is applied by setting the diagonal element $Y_{\nu_{\rm s}lm_l,\nu_{\rm s}lm_l}(r_{\rm in})=-ik_{\nu_{\rm s}lm_l}(r_{\rm in})$ while the off-diagonal elements of $\mathbf{Y}$ are set to zero \cite{Wang:2015}. Here, $k_{\nu_{\rm s}lm_l}(r_{\rm in})=\sqrt{E-V_{lm}^{\nu_{\rm s}}(r_{\rm in})}$ denotes the local wave number associated with the adiabatic channel $|\nu_{\rm s}lm_l\rangle$. We note that the inclusion of van-der-Waals interaction leads to $V_{lm}^{\nu_{\rm s}}(r_{\rm in})<E$ for all $|\nu_{\rm s}lm_l\rangle$ when $r_{\rm in}<15$ $a_0$. We found the van-der-Waals interaction plays a negligible role in the scattering process initialized in the $|\nu_{\rm s}^{\rm in}={++}\rangle$ state. However, for molecules prepared in certain other $|\nu_{\rm s}\rangle$ or $|\nu_{\rm a}\rangle$ states, the inclusion of the van-der-Waals interaction can be crucial. In the case of circular polarization without ellipticity $\xi=0$, we truncate the partial wave basis at $l<10$, which is sufficient for numerical convergence. Nevertheless, we find that truncation is generally required at $l<20$ when $\xi \neq 0$. 
\subsection{The FLR fit}
To extract the FLR parameters at a given $|\delta|/\Omega$, we fit the numerical obtained $\alpha_s$ using the following model: (i) $\alpha_s=\alpha_{s\rm bg}+\bar{\Delta}_1/(\Omega-\Omega_{\rm r_1})+b\Omega$ applied over the scan range $\Omega \in 2\pi \times [0,100]$ MHz; (ii) $\alpha_s=\alpha_{s\rm bg}+\bar{\Delta}_1/(\Omega-\Omega_{\rm r_1})+\bar{\Delta}_2/(\Omega-\Omega_{\rm r_2})$ applied over the extended scan range $\Omega \in 2\pi \times [0,800]$ MHz; and (iii) $\alpha_s=\alpha_{s\rm bg}+\bar{\Delta}_1/(\Omega-\Omega_{\rm r_1})+\bar{\Delta}_2/(\Omega-\Omega_{\rm r_2})+b\Omega$ applied also over $\Omega \in 2\pi \times [0,800]$ MHz. The fitted parameters for $|\delta|/\Omega=0$ are listed in Table \ref{tab:fit}. We found that the background scattering length $\alpha_{s\rm bg}$ and linear coefficient $b$ depend strongly on the scan range and the model. However, the resonance positions ($\Omega_{\rm r_1}$ and $\Omega_{\rm r_2}$ ) and effective width ($\bar{\Delta}_1$ and $\bar{\Delta}_2$) remain nearly unchanged across different fittings.
\end{document}